\documentclass[preprint,aps]{revtex4}
\usepackage{graphicx}
\usepackage{dcolumn}
\begin{document}


\title{ New physical effects on the decay $B_{s(d)} \to \gamma\gamma$  in the sequential fourth Generation model}


\author{Hong Chen$^{a}$}

\author{Wujun Huo$^{a,b}$}
\email{whuo@pku.org.cn}
\affiliation{$^a$College of Physics and Technology, Southwest
University, Chongqing 400715, China}
\affiliation{$^b$The Abdus Salam International Center for
Theoretical Physics, Strada Costiera 11, 34014  Trieste, Italy.}



\begin{abstract}
We study the contributions to  the branching ratios of $B_{s(d)}\to  \gamma \gamma$ decay
 in the sequential fourth generation model (SM4). We find that the
theoretical values of the branching ratios, ${\rm 
BR}(B_{s(d)}\to\gamma\gamma)$, including the contributions of $m_{t'}$ and the new
$4 \times 4$ CKM (CKM4)
matrix factors, $|V^{*}_{t's}V_{t'b}|$ and $|V^{*}_{t'd}V_{t'b}|$,
are much different from the minimal standard model (SM)
predictions. The new physics effects, especially contributed from the CKM4 matrix factors,
can provide more than one
order enhancement to the SM prediction. It is shown that the decay
$B_{s(d)}\to \gamma \gamma$ can test the new physics signals
from SM4.
\end{abstract}

\maketitle

\section{Introduction}

SM a very successful theory of the electroweak interactions. But it should not be the final
theory. Theoretically, it has too many unknown parameters to be put by
hand and can not unify the three gauge interactions. Also, SM has been faced to some troubles
 from the experiments.
We need the new physics beyond SM. Many new physics models have been proposed to resolve the difficulties of SM 
and to explain the experiments. Of course, they have to be tested in many
high energy experiments, such as the rare decays of mesons.

The startup of the LHC opens many new frontiers in precision flavour physics.
As is well known, the rare radiative decays of B mesons are particularly sensitive to the contributions
from  new physics. Both inclusive and exclusive processes have been researched  in the last 20 years.
For example,
 $B_{s(d)} \to \gamma \gamma$ has been studied extensively in the
 SM\cite{bsrr_SM} and new physics scenarios\cite{refs,huo_tc}.
The present experimental limit  on the decay $B_{s(d)} \to \gamma\gamma$ \cite{PRL100,ICHEP10}is
\begin{eqnarray}
\label{bsrr_exp}
 {\rm BR}(B_s \to \gamma\gamma) \leq 8.6 \times 10^{-6}\,\,(90\% C.L.) , \\
 {\rm BR}(B_d \to \gamma\gamma) \leq 3.2 \times 10^{-7}\,\,(90\% C.L.) .
\end{eqnarray}
Within the SM one finds\cite{bsrr_SM},
\begin{eqnarray}
\label{bsrr_SM}
 {\rm BR}(B_s \to \gamma\gamma) \simeq 1 \times  10^{-6}, \\
 {\rm BR}(B_d \to \gamma\gamma) \simeq 3 \times 10^{-8}.
 \end{eqnarray}
The upper bound of $B_{s(d)} \to \gamma\gamma$ is about ${\cal{O}}(1)$
larger than the SM values. We believe, with the continuous accumulate of the
experiment data, especially in the era of LHC and ILC, these branching ratios will be more and more
precise. They will leave less room for the new physics. That is to say, $B_{s(d)} \to \gamma\gamma$,
is very suitable to test the new physics models.

In ref. \cite{huo_tc}, we investigated the new physical effects on $B_{s}\to \gamma\gamma$
in the one generation Technicolor model (OGTM) and got some interesting results.
In this note, we consider the sequential fourth generation model\cite{sm4} to
 estimate the possible contributions to the decay  $B_{s(d)}\to \gamma\gamma$.  Recently, SM4  attracts an increasing
 interest and seems warming up. The electroweak precision data does not exclude completely existence of the fourth family 
and there are  many reasons to introduce an extra generation of heavy particles\cite{sm4}, (for a recent brief review 
on the 4th generation, see \cite{hou}).  Especially, LHC has the potential to discover or fully exclude  existence of a fourth generation of quarks up to 1 TeV\cite{hou},
even if they are too heavy to observe directly they will induce a large
signal in $g g \to Z Z$ that will be clearly visible at the LHC\cite{chan}. Maybe this model
will be firstly tested by the early LHC data.

 The sequential fourth generation model  is a simple and non-supersymmetric extension of the SM,  which
does not add any new dynamics to the SM,  with an additional up-type $t′$ and an down-type $d′$ quarks, a heavy charged 
lepton$\tau'$  and a heavy neutrino $\nu'$.

 The model retains all the properties of the SM. The $t′$ quark like the other up-type quarks contribute to the $b \to s$ 
transition at the loop level. Due to the additional fourth generation there will be mixing between the $t′$ quark the 
three down-type quarks of the standard model and
the resulting mixing matrix will become a $4 \times 4$ matrix,
\begin{equation}
V_{\rm CKM4} = \left (
\begin{array}{lcrr}
V_{ud} & V_{us} & V_{ub} & V_{ub'}\\
V_{cd} & V_{cs} & V_{cb} & V_{cb'}\\
V_{td} & V_{ts} & V_{tb} & V_{tb'}\\
V_{t'd}& V_{t's}& V_{t'b} &V_{t'b'}\\
\end{array} \right ) ,
\end{equation}
where $V_{qb'}$ and $V_{t'q}$ are the new matrix elements in the SM4. The parametrization of
this unitary matrix requires six mixing angles and three phases\cite{CKM4}.

\section{Branching ratios of $B_{s(d)} \to \gamma \gamma$}

At quark level, $b \to s \gamma $, $b \to s \gamma \gamma$ and the
exclusive decays $B_{s} \to \gamma \gamma$ have a close relation.
Up to the corrections of order $1/m_W^2$,
  the effective Hamiltonian for
 $b\to s\gamma\gamma$ at scales   $\mu_b={\cal O}(m_b)$ is identical to the one for
 $B \to X_s \gamma$ transition \cite{bsrr_SM} and takes the form
 \begin{widetext}
    \begin{eqnarray}
        {\cal H}_{\rm eff} =
           \frac{G_{\rm F}}{\sqrt{2}} V_{ts}^* V_{tb}
           \left[ \sum_{i=1}^6 C_i(\mu_b) Q_i +
           C_{7\gamma}(\mu_b) Q_{7\gamma }
           +C_{8G}(\mu_b) Q_{8G} \right], \label{eq:heff}
    \end{eqnarray}
   \end{widetext}
 here $Q_1 \dots
 Q_6$ are the usual four-fermion operators whose explicit form is given below.
  The last two operators in the Eq.(\ref{eq:heff}),
 characteristic for this decay, are the {\it magnetic--penguin} operators.
 The complete list of operators is given as follows
\begin{eqnarray}
Q_1&=&(\bar{c}_{L\beta} \gamma^{\mu} b_{L\alpha})
            (\bar{s}_{L\alpha} \gamma_{\mu} c_{L\beta}),\\
Q_2&=&(\bar{c}_{L\alpha} \gamma^{\mu} b_{L\alpha})
            (\bar{s}_{L\beta} \gamma_{\mu} c_{L\beta}),\\
Q_3&=&(\bar{s}_{L\alpha} \gamma^{\mu} b_{L\alpha})
\sum_{q=u,d,s,c,b}(\bar{q}_{L\beta} \gamma_{\mu} q_{L\beta}),\\
Q_4&=&(\bar{s}_{L\alpha} \gamma^{\mu} b_{L\beta})
\sum_{q=u,d,s,c,b}(\bar{q}_{L\beta} \gamma_{\mu} q_{L\alpha}),\\
Q_5&=&(\bar{s}_{L\alpha} \gamma^{\mu} b_{L\alpha})
 \sum_{q=u,d,s,c,b}(\bar{q}_{R \beta} \gamma_{\mu} q_{R\beta}),\\
Q_6&=&(\bar{s}_{L\alpha} \gamma^{\mu} b_{L\beta})
 \sum_{q=u,d,s,c,b}(\bar{q}_{R\beta} \gamma_{\mu} q_{R\alpha}),\\
Q_7&=&(e/16\pi^2) m_b \bar{s}_L \sigma^{\mu\nu}
            b_{R} F_{\mu\nu},\\
Q_8&=&(g/16\pi^2) m_b \bar{s}_{L} \sigma^{\mu\nu}
            T^a b_{R} G_{\mu\nu}^a. \label{eq:o8g}
\end{eqnarray}
 where $\alpha$ and $\beta$ are color indices, $\alpha=1, . . . ,8$
labels $SU(3)_C$ generators, $e$ and $g$ refer to the
electromagnetic and strong coupling constants, while $F_{\mu\nu}$
and $G^{a}_{\mu\nu}$ denote the QED and QCD field strength tensors,
respectively. It is the magnetic $\gamma$-penguin operator $Q_7$,
which plays the crucial role in this
 decay. The effective Hamiltonian for $b\to d\gamma\gamma$ is
 obtained from Eqs.(\ref{eq:heff}-\ref{eq:o8g}) by the replacement $s\to d$.

\begin{figure}[th]
\includegraphics{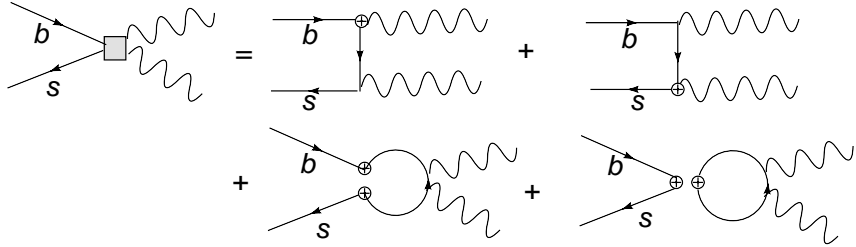}
\caption{Examples of Feynman diagrams that contribute to the matrix
element. \label{fig1}}
\end{figure}

The Feynman diagrams that contribute to the matrix element as the
following, see Fig. \ref{fig1},

Within the SM, at scale $m_W$, the Wilson coefficients $C_i(m_W)$
at the leading order (LO) approximation have been given for example
in \cite{buras98},
%
 \begin{eqnarray}
C_i(m_W ) &= & 0 \ \ \ (i=1,3,4,5,6), \,\,C_2(m_W)  =  1, \label{eq:c2mw} \\
C_7(m_W) &=& \frac{8x_t^3+ 5x_t^2-7 x_t   }{ 24(1-x_t)^3}
    - \frac{ 2 x_t^2-3 x_t^3 }{ 4(1-x_t)^4}\log[x_t], \label{eq:c7mw}\\
C_8(m_W) &=& \frac{x_t^3- 5x_t^2-2 x_t   }{ 8(1-x_t)^3} - \frac{ 3
x_t^2 }{ 4(1-x_t)^4}\log[x_t], \label{eq:c8mw}
 \end{eqnarray} 
 %
 where $x_t=m_t^2/m_W^2$.

By using QCD renormalization group equations\cite{buras98}, it is
straightforward to run Wilson coefficients $C_i(m_W)$ from the scale
$\mu ={\cal O}( m_w)$ down to the lower scale $\mu ={\cal O}(m_b)$.
The leading order results for the Wilson coefficients $C_7(\mu)$
with $\mu \approx m_b$ are of the form \cite{buras98}
%
\begin{eqnarray}
C_7(\mu) &=& \eta^{16/23} C_7(m_W) +\frac{8}{3} \left (
\eta^{14/23} - \eta^{16/23} \right ) 
  C_8(m_W) + \sum_{i=1}^8 h_i \eta^{a_i}\, , \label{eq:c7mu}
\end{eqnarray}
%
 where
$\eta=\alpha_s(m_W)/\alpha_s(\mu)$,
%
\begin{eqnarray}
a_i&=& ( 14/23,16/23,6/23,-12/23,\\ \nonumber
  &&0.4086, -0.4230, -0.8994,0.1456)\, , \label{eq:ai}\\
h_i &=& ( 2.2996,-1.0880,-3/7,-1/14,\\ \nonumber
  &&-0.6494,-0.0380,-0.0185,-0.0057)\,   \label{eq:hi}.
\end{eqnarray}
%

\begin{figure}[th]
{\includegraphics{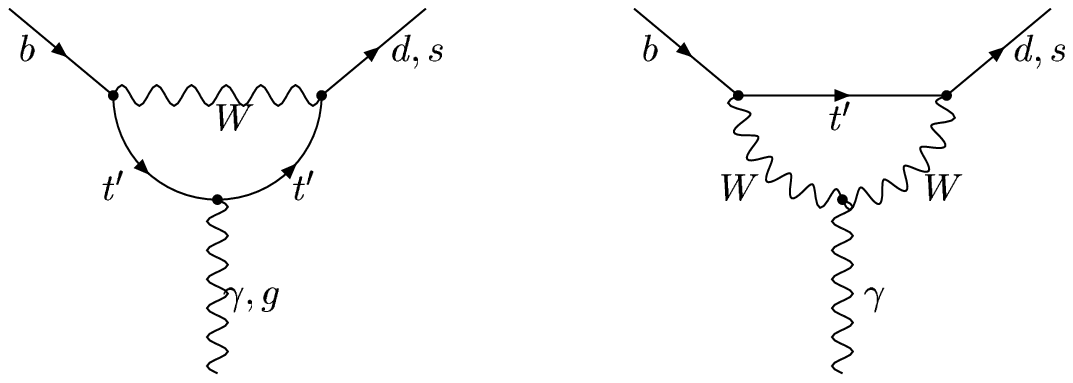}}
\caption
{Magnetic photon and gluon penguin diagrams  with the fourth
generation $t'$ quark.} \label{fig2}
\end{figure}

 In the sequential 4th generation model,  there exists  an additional contribution to $b \to  s\gamma$
induced by  the 4th generation up quark $t'$, which produce the
penguin diagrams, see Fig. (\ref{fig2}),

The new  Wilson coefficients contributed by $t'$ are
same as their counterparts in Eq. (\ref{eq:c7mw}) and (\ref{eq:c8mw})
 except for exchanging  $t'$ quark not $t$ quark.

At the mass scale of $\mu_b$, the Wilson
coefficients of the dipole operators are given by
   \begin{equation}
      C^{\rm eff}_{7,8}(\mu_b)=C^{\rm (SM)\rm eff}_{7,8}(\mu_b)
      +\frac{V^{*}_{t^{'}s}V_{t^{'}b}}{V^{*}_{ts}V_{tb}}C^{(4)
      {\rm eff}}_{7,8}(\mu_b),
    \end{equation}
 where
$V^{*}_{t^{'}s}$ and $V_{t^{'}b}$ are two elements of the $4\times
4$ CKM matrix. We recall here that the CKM coefficient
corresponding to the $t$ quark contribution, i.e., $V_{ts}^*V_{tb}$,
is factorized in the effective Hamiltonian as given in Eq. (\ref{eq:heff}).

To calculate $B_{s(d)} \to \gamma\gamma$ , one may follow a perturbative
QCD approach which includes a proof of factorization, showing that
soft gluon effects can be factorized into $B_{s(d)}$ meson wave
function; and a systematic way of resuming large logarithms due to
hard gluons with energies between 1Gev and $m_{b}$. In order to
calculate the matrix element of Eq(1) for the $B_{s(d)} \to \gamma\gamma$
, we can work in the weak binding approximation and assume that both
the $b$ and the $s(d)$ quarks are at rest in the $B_{s(d)} $ meson, and the
$b$ quarks carries most of the meson energy, and its four velocity
can be treated as equal to that of $B_{s(d)} $. Hence one may write $b$
quark momentum as $p_{b}=m_{b}v$ where is the common four velocity
of $b$ and $B_{s(d)}$. We have
%
\begin{eqnarray}
p_{b}\cdot k_1&=&m_bv\cdot k_1=\frac{1}{2}m_bm_{B_{s(d)}}=p_{b}\cdot k_2,\nonumber \\
p_{s(d)}\cdot k_1&=&(p-k_1-k_2)\cdot k_1\nonumber \\
&=& - \frac{1}{2}m_{B_{s(d)}}(m_{B_{s(d)}}-m_b) \nonumber \\
&=&p_{s(d)}\cdot k_2.
\end{eqnarray}
%

We compute the amplitude of $B_{s(d)} \to \gamma\gamma$ using the
following relations
\begin{eqnarray}
\left\langle 0\vert \bar{s}(\bar{d})\gamma_{\mu}\gamma_5 b\vert B_{s(d)}(P)
\right\rangle
&=& -if_{B_{s(d)}}P_{\mu},\nonumber \\
\left\langle 0\vert \bar{s}(\bar{d})\gamma_5 b\vert B_{s(d)}(P) \right\rangle &=&
if_{B_{s(d)}}M_B,
\end{eqnarray}
where $f_{B_{s,d}}$ is the $B_{s(d)}$ meson decay constant.

The total amplitude is now separated into a CP-even and a CP-odd
part
\begin{equation}
T(B_{s(d)}\to \gamma\gamma)=M^+F_{\mu\nu}F^{\mu\nu}
+iM^-F_{\mu\nu}\tilde{F}^{\mu\nu}.
\end{equation}
We find that
\begin{widetext}
\begin{eqnarray}
M^+&=&-\frac{4{\sqrt 2}\alpha G_F}
{9\pi}f_{B_{s(d)}}m_{B_{s(d)}}V_{ts(d)}^*V_{tb}  
\left(B m_b K(m_b^2) 
+\frac{3C_7}{8\bar{\Lambda}}
\right),\\
M^-&=&\frac{4{\sqrt 2}\alpha G_F}
{9\pi}f_{B_{s(d)}}m_{B_{s(d)}}V_{ts(d)}^*V_{tb} 
 \left(\sum_q 
m_{B_{s(d)}}A_q J(m_q^2)+ m_b B L(m_b^2) 
+\frac{3C_7}{ 8\bar{\Lambda}}
\right),
\end{eqnarray}
\end{widetext}
with $B= -(3C_6+C_5)/4$, $ \bar{\Lambda}=m_{B_{s(d)}}-m_b$, and
\begin{eqnarray}
A_u &=&(C_3-C_5)N_c+(C_4-C_6), \nonumber \\
A_d &=&\frac{1}{4}\left[(C_3-C_5)N_c+(C_4-C_6)\right], \nonumber \\
A_c &=&(C_1+C_3-C_5)N_c+(C_2+C_4-C_6), \nonumber \\
A_s & =& \frac{1}{4} \left[(C_3+C_4-C_5)N_c+(C_3+C_4-C_6)\right],\nonumber \\
A_b &=& A_s.
\end{eqnarray}
The functions $J(m^2)$, $K(m^2)$ and $L(m^2)$  are defined by
\begin{eqnarray}
J(m^2)&=& I_{11}(m^2),\nonumber \\
K(m^2)&=& 4 I_{11} (m^2)- I_{00}(m^2) ,\nonumber \\
L(m^2)&=&I_{00}(m^2).
\end{eqnarray}
with
\begin{equation}
I_{pq}(m^2)=\int_{0}^{1}{dx}\int_{0}^{1-x}{dy}\frac{x^{p}y^{q}}{m^{2}-2xyk_{1}\cdot
k_{2}-i\varepsilon}.
\end{equation}

The decay width for $B_{s(d)}\to \gamma\gamma$ is simply
\begin{equation}
\Gamma(B_{s(d)}\to \gamma\gamma)=\frac{m_{B_{s(d)}}^3}{ 16\pi}({\vert M^+\vert
}^2+{\vert M^-\vert }^2).
\end{equation}

\section{Numerical analysis and summary  }

In the numerical calculations we use as input parameters $\alpha_s (m_Z ) = 0.118$,
, $\alpha_s(m_b)= 0.223$, $m_W = 80.22 {\rm GeV}$, $m_c = 1.27 {\rm GeV}$, $m_b = 4.19 {\rm GeV}$, 
$m_t = 172{\rm GeV} $, $\tau_{B_s} = 1.49ps$,  $f_{B_s}= 230{\rm MeV}$,
 $\lambda_{B_s}= \lambda_{B_d}= 350 {\rm MeV}$,  $m_{B_s} = 5.37{\rm GeV}$,  $\tau_{B_d} =1.55ps$,   
 $f_{B_d} = 200{\rm MeV}$ and $m_{B_d}=5.28{\rm GeV}$, respectively.

For the mass limit of $t'$,  CDF
gives  $m_{t'} > 256 {\rm GeV}$  for
the $t' \to qW$ final state \cite{mt'}.
The experimental upper bounds for the fourth family quark CKM  matrix elements
are $|V_{t' d }| < 0.063$, $|V_{t' s} | < 0.46$, $|V_{t' b} | < 0.47$ \cite{exp_ckm4}.
By taking the CKM unitarity conditions,
$
 \sum\limits_{i}V_{is(d)}^{*}V_{ib}=0,\,\,(i=u,c,t,t'),
$
and the present measurement of $3 \times 3$ CKM matrix\cite{PDG},
We obtain the bounds for the  CKM4 matrix elements in SM4,
\begin{eqnarray}
|V^{*}_{t'd}V_{t'b}|& < & (1.83-2.03)\times 10^{-2}, \\
|V^{*}_{t's}V_{t'b}|& < & (6.97-7.75)\times 10^{-2}.
\end{eqnarray}

\begin{figure}[th]
{\includegraphics{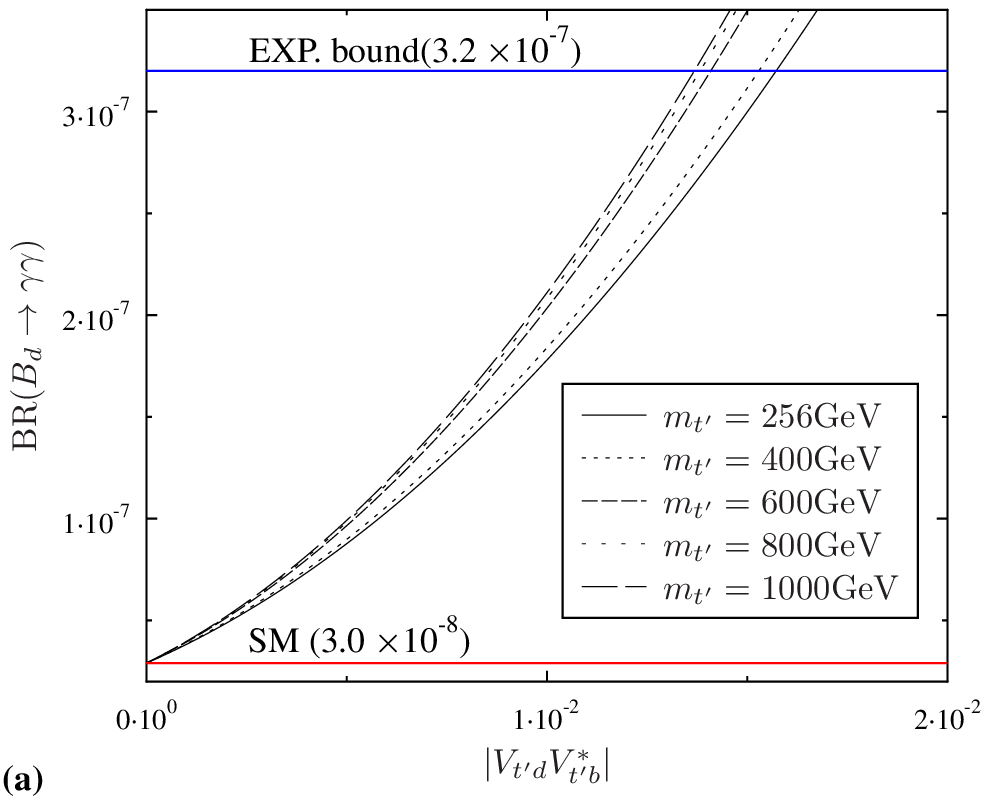}}
{\includegraphics{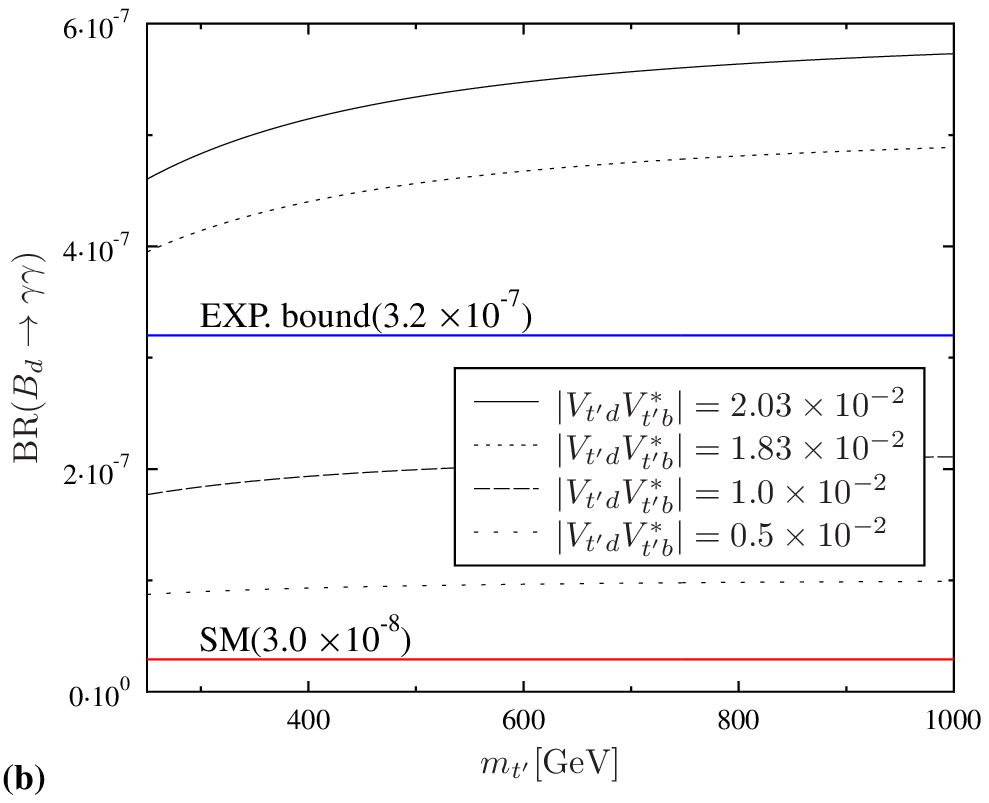}}
\caption
{The Branching ratio of $B_d \to \gamma \gamma $ versus (a) CKM4 matrix factor $|V^*_{t'd}V_{t'b}|$ for
 different values of $m_{t'}$; (b) the mass of $t'$
for different values of $|V^*_{t'd}V_{t'b}|$.} \label{fig3}
\end{figure}

Fig. \ref{fig3}a  shows the dependence of
$ {\rm BR}(B_d \to \gamma\gamma)$ with the CKM4 matrix factor $|V^{*}_{t'd}V_{t'b}|$
for different values of $m_{t'}$. We can see that the new
physics contributions can lead to appreciable changes
of the SM predictions which may be enhanced by
about more than one orders of magnitude in a reason-
able mass range for $t'$. The new physics effects is very
sensitive to the value of $|V^{*}_{t'd}V_{t'b}|$ and becomes tiny
as $|V^{*}_{t'd}V_{t'b}| < 0.5 \times 10^{-2}$. It mains that
$B_s \to \gamma \gamma$ can give the great limit room for the CKM4 matrix elements and
give the strong constrict for the contributions to CP violation in SM4. But
from Fig. \ref{fig3}a, the new physics effects is not sensitive
to the mass of $t'$. This can be seen more clearly in Fig. \ref{fig3}b, which
shows the mass dependence of
$ {\rm BR}(B_d \to \gamma\gamma)$ with $t'$
for different values of $m_{t'}$. From Fig. \ref{fig3}b, we can see that the
new physics effects  become bigger with increasing mass of the $t'$. This  is similar to the case of 
top quark to the rare $B$ meson decays
in SM, which the main contributions come from the heavy quark.


\begin{figure}[th]
{\includegraphics{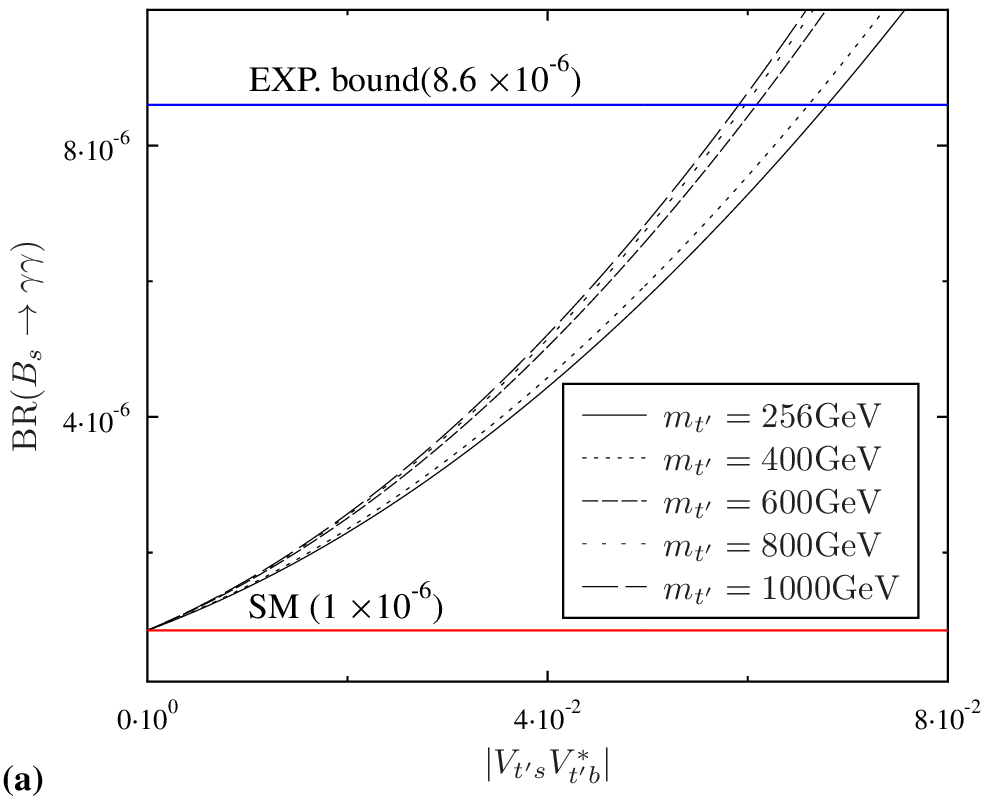}}
{\includegraphics{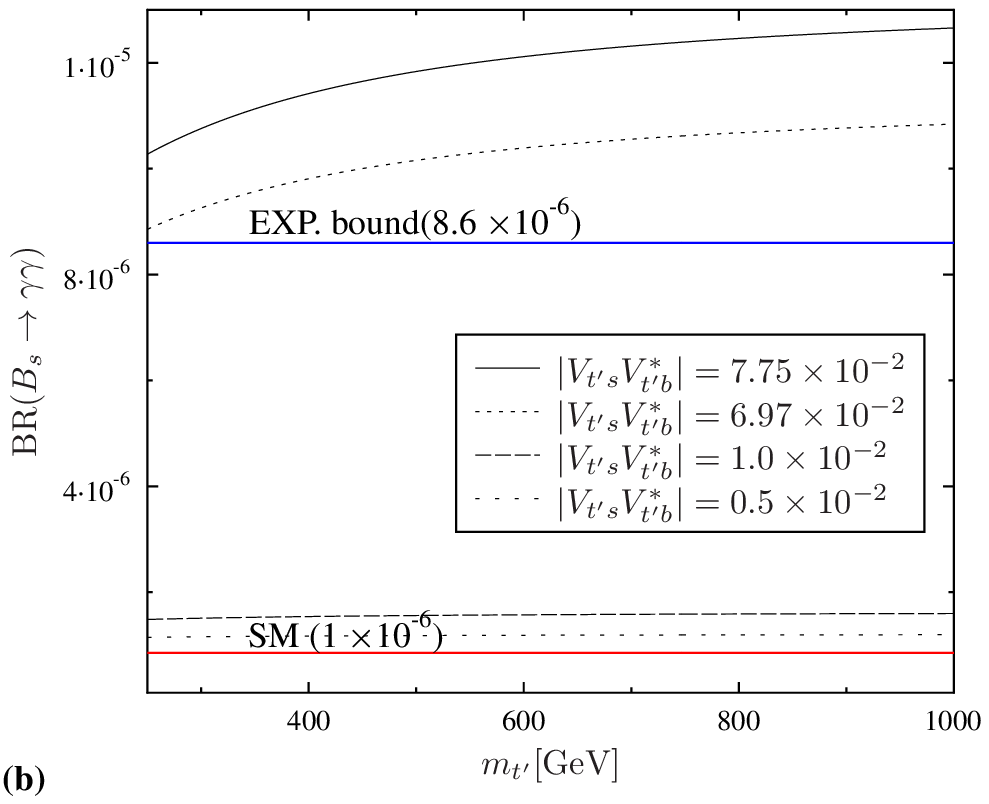}}
\caption
{The Branching ratio of $B_s \to \gamma \gamma $ versus (a) CKM4 matrix factor  $|V^*_{t's}V_{t'b}|$ 
for different values of $m_{t'}$; (b) the mass of $t'$
for different values of $|V^*_{t's}V_{t'b}|$.} \label{fig4}
\end{figure}
%

Figs. \ref{fig4} and  show the dependence of the decay
${\rm BR}(B_s \to \gamma \gamma) $ with the CKM4 matrix factor $|V^{*}_{t's}V_{t'b}|$
for different values of $m_{t'}$. We can get the similar analysis but
the new physics effects is much more
sensitive to the value of the CKM4 matrix factor, $|V^{*}_{t's}V_{t'b}|$.
For both of these decays, the CKM4 matrix elements provides the dominant
new physics contribution.

As a conclusion, the new physics contribution to
the rare decay of $B_{s(d)} \to  \gamma \gamma$ in the sequential fourth
 can enhance  rather large in magnitude, and may be detected in the near
future precision experiments.

\begin{acknowledgments}
W. Huo  thanks the SWU
for the invitation and accommodation. This work was supported in parts by 
SRF for ROCS, SEM, P. R. Chna
 (Grant No. 6807020000).
\end{acknowledgments}

\end{document}